# *synapse*: interactive support on photoemission spectroscopy measurement and analysis for non-expert users

Takuma Masuda,[a] Masaki Kobayashi,[a,b]* and Koji Yatani[a†]

[a]*Department of Electrical Engineering and Information Systems, The University of Tokyo, 7-3-1 Hongo, Bunkyo-ku, Tokyo 113-8656, Japan*

[b]*Center for Spintronics Research Network, The University of Tokyo, 7-3-1 Hongo, Bunkyo-ku, Tokyo 113-8656, Japan.*

*Email: masaki.kobayashi@ee.t.u-tokyo.ac.jp

†Email: koji@iis-lab.org

Photoemission spectroscopy, an experimental method based on the photoelectric effect, is now an indispensable technique used in various fields such as materials science, life science, medicine, and nanotechnology. However, part of the experimental process of photoemission spectroscopy relies on experience and intuition, which is difficult to understand for novice users. In particular, photoemission spectroscopy experiments using high-brilliance synchrotron radiation as a light source are not easy for novice users because the measurements must be performed quickly and accurately as scheduled within a limited experimental period. In addition, research on the application of information science methods to quantum data measurement, such as photoemission spectroscopy, is mainly aimed at the development of analysis methods, and few attempts have been made to clarify the problems faced by users who lack experience. In this study, we identified the problems faced by novice users of photoemission spectroscopy, implemented a native application with functions to solve these problems, and evaluated it qualitatively and quantitatively. This paper describes the contents of the field study and interview survey, the functional design and implementation of the application based on our field study and interview survey, and the results and discussion of the evaluation experiment.

**1. Introduction**

Photoemission spectroscopy is now indispensable in various fields such as materials science, life science, medicine, and nanotechnology (Reinert1 & Hüfner, 2005; Winter & Faubel, 2006; Peles & Simon, 2009; Vilmercati *et al*., 2009; Kobayashi, 2009; Papp & Steinrück, 2013; Lv *et al*., 2019). However, part of the experimental process of photoemission spectroscopy relies on experience and intuition, making it difficult for novice users to understand. For example, while experimenters of photoemission spectroscopy repeat measurements under the same conditions and aggregate the results to increase the signal-to-noise ratio (SNR) of a spectrum, they decide how many iterations they need visually and empirically. Given the increasing number of users of photoemission spectroscopy across the boundaries between academia and industry, it should be easy to understand even for non-skilled users.

In photoemission spectroscopy, synchrotron radiation is frequently used as a light source. The use of synchrotron radiation as a light source increases SNR in short measuring time due to the highly brilliant X-rays and improves measurement accuracy dramatically. In addition, users require fewer iterations for the measurement, and the measurement time is expected to be reduced. For these reasons, the light source is often used for photoemission spectroscopy.

However, while synchrotron radiation light sources are becoming increasingly popular, photoemission spectroscopy experiments using synchrotron radiation are particularly burdensome for unfamiliar users. Public use of these facilities is generally limited to a few times a year for a few hours to a few days each time, during which experimental teams must repeat scheduled measurements. In other words, the users need to perform the scheduled experiment without failure in the limited time available. In addition, users must analyze the spectra and consider the next measurement in a relatively short measurement time, which requires more accurate and quicker processing than in a laboratory experiment. Therefore, it is especially burdensome for users who are unfamiliar with the facility, equipment, and experimental techniques.

In this study, we propose *synapse* (SYNchrotron radiation App for Photoemission Spectroscopy Experiment), a measurement and analysis support system for novice users of photoemission spectroscopy. To build the system, we first conducted interviews with students and experts to clarify issues. Next, we formulated three features that should be implemented. Finally, we built an application with those functions and evaluated it quantitatively and qualitatively. In this paper, the related works will be described first, followed by functions, implementation, evaluations, discussion, and conclusion.

## 2. Related Works

### 2.1. Use of Informatics in Quantum Beam Measurement

Various measurements have been made using beams of photons, neutrons, and other quanta. In recent years, informatics methods are becoming prevalent in such quantum beam measurements including photoemission spectroscopy, motivated by the need to reduce measurement time (Ueno *et al.*, 2018; Ueno *et al.*, 2021), to help users calculate (Devereaux *et al.*, 2021), or to carry out simulation (Rakitin *et al.*, 2018; Suzuki *et al.*, 2019). For example, Ueno *et al.* proposed an adaptive experimental design method and developed a machine learning method for fitting curves to fit spectral plots, reducing the number of plots required for measurement (Ueno *et al.*, 2018). They also determined the convergence of iterative measurements necessary to increase SNR and developed a specific method to reduce the number of unnecessary iterations (Ueno *et al.*, 2021). In addition, there are more examples in quantum beam measurements such as X-ray absorption spectroscopy (XAS) (Suzuki *et al.*, 2019; Timoshenko & Frenkel, 2019) and X-Ray Diffraction (XRD) (Hou *et al.*, 2019), some of which leverage machine learning. These attempts are called measurement informatics (Ueno *et al.*, 2018) or materials informatics (Rajan, 2005), which is one of the areas that has been getting a lot of attention lately. It has been pointed out that for the application of informatics methods to materials science, the design of the platform is as important as the construction of the database and the application of machine learning (Takahashi & Tanaka, 2016). However, few studies have identified what novices struggle with before design. Therefore, in this study, we first interviewed students and experts who are in a position to educate them to identify what novice users of photoelectron spectroscopy have trouble with.

## 2.2. Support by User Interface

There are many attempts to assist users by building interfaces. For example, *WIFIP* (Sallaz-Damaz & Ferrer, 2017) provides a remote control interface for evolving devices. Also, *Xi-cam* (Pandolfi *et al*., 2018) enables users to organize, view, and analyze image data with plugins. However, as mentioned earlier, there are few examples to identify and solve the problems that users of photoemission spectroscopy face in measurement and analysis. In particular, many existing interfaces focus on simply displaying spectra, processing and analyzing data, or connecting hardware and software.

In addition, Web-based user interfaces, such as *Web Ice* (GonzÅLalez *et al*., 2008), *WIFIP* (Sallaz-Damaz & Ferrer, 2017), *Sirepo* (Rakitin *et al*., 2018), and *Daiquiri* (Fisher *et al*., 2021) are becoming more and more common these days. This is presumably because using a technology stack like that used for the Web saves time and effort in building applications. On the other hand, applications that run in a Web browser may use a command-line user interface (CUI) to start up, making them difficult for users to use. Therefore, in this study, we decided to use Electron, a framework that can build native applications without OS dependence, to benefit from a Web-based technology stack and complete everything on the graphical user interface (GUI).

## 3. Support by User Interface

### 3.1. Interview survey

First of all, we conducted an interview survey to identify challenges novice users face in the experimental process of photoemission spectroscopy. Users who had experience with photoemission spectroscopy experiments participated in this survey, including experts and students. The survey results revealed two major problems of photoemission spectroscopy experiments: peak identification and SNR determination.

The first problem is peak identification. In photoemission spectroscopy, spectral peaks are an important piece of information, because the positions of the peaks tell what kinds of elements samples have in themselves. Peaks mainly derive from photoelectrons of the sample, electrons emitted from the inner shells of the elements due to the photoelectric effect. Other possible sources include noise, photoelectrons originating from carbon dioxide attached to the sample, and Auger electrons emitted as a side effect of photoelectron excitation. Participants of the interview survey pointed out that it is difficult to distinguish between peaks of photoemission spectra that come from constituent elements of the samples and other peaks. Also, in some cases, errors in measurement were noticed later, or the errors were not noticed in the first place. In particular, this problem is serious in the measurements using synchrotron radiation because if you miss the measurement, you cannot take measurements again for almost half a year.

The other problem is SNR. The SNR of spectra is increased by repeating measurements under the same conditions and aggregating them. However, when to stop the iteration is determined visually and empirically from the spectra. In particular, the use of synchrotron radiation as a light source makes it more difficult for novice users to carry out photoemission spectroscopy experiments because if they obtain spectra that cannot reach the necessary levels of SNR, they have to conduct the measurement again half a year later.

**3.2. Function design**

Given the two major problems found by the interview survey, we decided to implement the following four functions in *synapse*:

1. Function to formulate and visualize the SNR

2. Function to display the binding energy of the constituent elements of the sample in the spectrum

3. Function to manage meta information and data in the form of log notes

First, we propose a method for calculating SNR. Given the fact that SNR is currently judged visually and empirically from spectra, it would be better to evaluate it quantitatively. Also, the SNR is not only calculated but also visualized so that users can quickly grasp it.

In addition, we put into the spectrum the binding energy of the elements that possibly consist of the sample. Currently, users have to reference data tables of binding energy so that they can identify the origin of the peaks. Therefore, they have to search data from a large database (often in the form of physical books), which makes it difficult for unfamiliar users to identify peaks. With biding energy on the spectra, it would be easier to figure out the origin.

We also organize the measurement information of the experiments following the form of existing log notes, in which experimenters record the values of parameters used and experimental results, such as the sample position, the intensity of light, and the presets used in the measurement. By adopting this log note format, it is expected that users will be able to use the application according to their existing user experience.

**4. Implementation**

**4.1. Architecture**

In this section, we discuss how to build the entire application, not each function. This section is divided into the data input/output part, the logic part, and the interface part, and is explained in detail.

In many cases, there is no Application Programming Interface (API) or Software Development Kit (SDK) provided for photoemission spectroscopy instruments. Even if it is available, it is often limited to LabVIEW[1] (Kalkman, 1995). Therefore, we implemented the system so that data files are loaded and processed after measurement. This is the same as when processing is done by analysis software such as Igor[2], so the existing user experience is not compromised.

The core logic of *synapse* was basically implemented in TypeScript, which is a strongly typed programming language that builds on JavaScript. We use various packages provided by the npm community to save time and effort in implementation. For example, *synapse* is built on Electron, a framework enabling developers to build crossplatform desktop apps, Next.js[3], a React[4] framework facilitating static site generation (SSG) and server-side rendering (SSR), and MUI, a React[5] component library with material design[6]. The architecture of *synapse* is shown in Fig. 1(a). In the main process, *synapse* communicates with file API, which enables users to read or write data files. Via IPC, the main process communicates with the renderer process, in which the application shows the user interface.

---

[1] Graphical programming environments developed by National Instruments Corporation (https://www.ni.com/ja-jp/shop/labview.html).

[2] Software for analyzing spectra provided by Wavemetrics (https://www.hulinks.co.jp/software/da_visual/igor/section02).

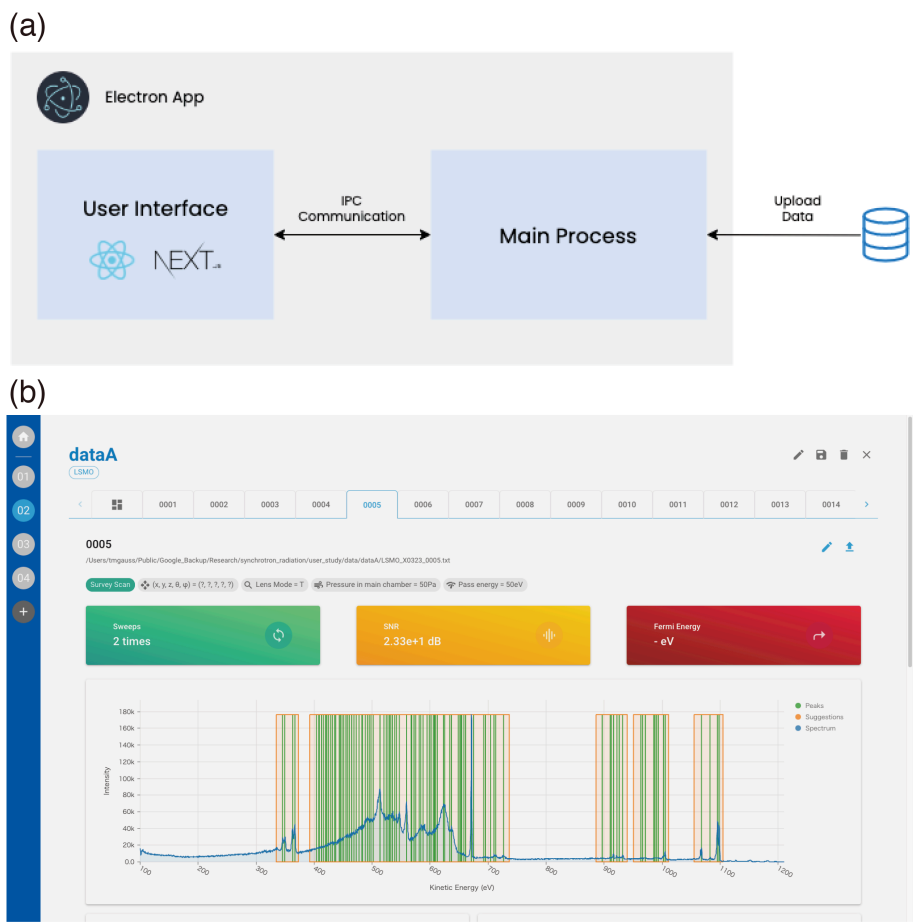

Fig. 1. Graphical user interface *synapse*. (a) Architecture. (b) User interface.

---

[3]https://nextjs.org/

[4]https://ja.reactjs.org/

[5]A front-end package for building user interfaces based on UI components maintained by Meta (formerly Facebook)

[6]The design scheme developed by Google.

The application interface was also basically implemented in TypeScript. We use Next.js to improve the performance of the application and MUI to save time to develop React components from scratch. The interface is shown in Fig. 1(b). The following is a brief description of how to use the system. First, create a project and enter the names and constituent elements of the samples to be used. Then, add measurement data along with metadata about the equipment, sample location, light source, etc. After that, the dashboard is displayed, in the center of which the spectrum is displayed with the binding energies of the constituent elements. Above that, accumulations, SNR, and Fermi levels are visualized.

### 4.2. Formula of SNR

Currently, there is no prevalent method for calculating SNR in photoemission spectroscopy. Therefore, we first summarized the requirements that SNR must meet.

- Dimensionless
- Able to be compared across spectra
- Monotonically increasing according to the number of iterations

To achieve these requirements, we decided to compute SNR using the Savitzky-Golay filter (Savitzky & Golay, 1964), a smoothing filter that can reduce the noise.

Generally, a smoothing filter modifies each data point using its adjacent points. One of the simplest smoothing filters is the moving average filter, which converts a certain data point into an average of its adjacent points. Let $x_i$ be the $i$-th data point and $x'_i$ the point converted from $x_i$ by a moving average filter. Then, $x'_i$ can be described as follows.

$$x'_i = \frac{1}{2N + 1} \sum_{j=i-N}^{j=i+N} x_j$$

$2N + 1$ is a window length, which means that the point, the N points to the left of it, and the $N$ points to the right of it are used. Savitzky-Golay filter is very similar to the moving average filter. While a moving average filter replaces a data point with a mean of its adjacent points, a Savitzky-Golay filter replaces a data point with a central point of the polynomial curve that fits its adjacent points by the method of linear least squares. Its parameters are window length and polynomial order.

Let $I_i$ ($i = 0, 1, 2, \ldots, N - 1$) be the intensity of the $i$-th data point. Let $I_{i,\text{net}}$ ($i = 0, 1, 2, \ldots, N-1$) be the discrete data of $I_i$ to which the Savitzky-Golay filter is applied, respectively. Based on the results of Furukawa's experiment (Furukawa *et al.*, 2016), we let the window length be 11, while based on our preliminary experiment, we temporally let the polynomial order be 4. Also, define the noise as follows.

$$I_{i,\text{noise}} = I_i - I_{i,\text{net}}$$

Using these, we define SNR as follows.

$$\text{SNR} = 10 \log_{10} \frac{E_i[I_{i,\text{net}}^2]}{E_i[I_{i,\text{noise}}^2]}$$

Note that $E_i[I_{i,j}^2]$ ($j$ = net, noise) represents the average of $I_{i,j}^2$ and the unit is dB, which is of course a dimensionless quantity.

## 5. Evaluation

To evaluate *synapse*, we conducted two types of evaluation experiments, a quantitative evaluation, and a qualitative evaluation. First, we quantitatively evaluated the formula of SNR, while qualitatively investigating the user experience of *synapse*. This chapter describes the evaluations.

### 5.1. Quantitative evaluation of formulation of SNR

We applied the methods described in section 4.2 to actual data and plotted intensity and SNR according to kinetic energy and sweeps, respectively. We used a window width of 11 and a polynomial degree of 4 for the parameters of the filter, referring to the work of Furukawa *et al* (Furukawa *et al*., 2016). The results are shown in Fig. 2. Judging visually, it is clear that the SNR of the Sr 3*d* spectrum is higher than that of the Sr 4*d* spectrum, as shown in Figs. 2(a) and 2(c). This is consistent with the results of comparing the SNRs of the rightmost point on the right side of Fig. 2(b) and that of Fig. 2(d). Furthermore, it is confirmed that the SNR increases almost monotonically with the number of accumulations, as shown in Figs. 2(b) and 2(d).

### 5.2. Qualitative evaluation of user experience

We conducted a user study in which one expert and one student were asked to use the application and interviewed about their experience. In the experiment of the expert, he made actual measurements and analyzed the data at a synchrotron radiation facility. On the other hand, the experiment of the student was a simulated analysis using pre-measured data. Afterward, we conducted interviews, in which the questions mainly focused on this point: "Is it easier to stop the repetition of measurement (Function 1)".

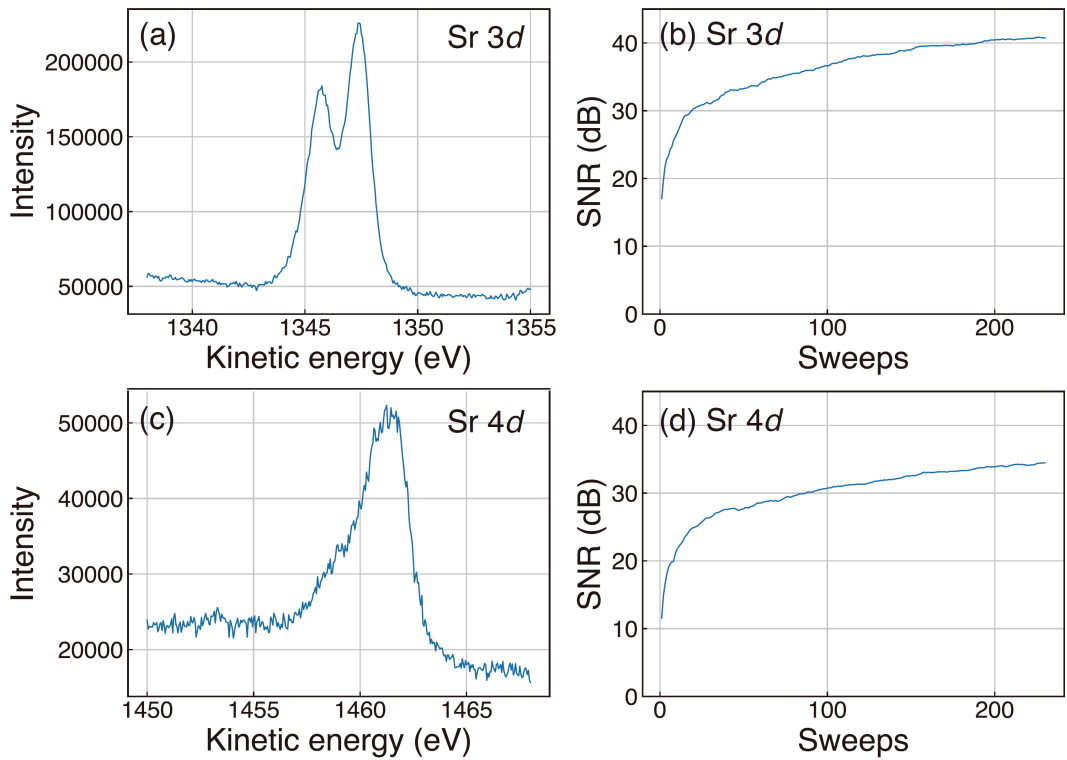

Fig. 2. Evaluation of SNR. (a) Data on the spectrum of the $3d$ orbital of Sr. The number of accumulations is 230. (b) SNR of $3d$ orbitals of Sr. The horizontal axis is the number of accumulations, and the vertical axis is the SNR. (c) Data on the spectrum of the $4d$ orbital of Sr. The number of accumulations is 230. (d) SNR of $4d$ orbitals of Sr. The horizontal axis is the number of accumulations, and the vertical axis is the SNR.

The format of the interview was a semi-structured interview based on the following questions prepared in advance, with further questions asked according to the participants' answers.

• Did you require less effort to perform the analysis using this application? Evaluate the performance on a scale of 1 to 5 and tell us why you chose that option.

• How has your performance changed using this application (5-point scale)? Please tell us why you chose that option.

• Compared to the existing interface, is the new interface easier to use (on a 5-point

scale)? Please tell us why you chose that option.

• Would you recommend this application to someone you know (on a 5-point scale)? And why did you choose that option? What is the reason?

• In addition to the previous answers, what aspects of the system have you found easier to use?

• In addition to the previous answers, what improvements should be made?

• Did you find the measurement after the survey scan to be clear? Please answer with the reason.

• Has the SNR been clarified? Please answer along with your reasons.

As a result, the opinions were very positive about the ability to quantitatively calculate and visualize SNRs (Function 1). The expert said that the function is the best part of all the functions of *synapse* because people without experience do not know how much experimental data they need to accumulate. Also, the student added that if SNR is expressed in numbers, it gives us an indicator of how much iterations users should take.

Other than Function 1, the opinions were obtained about the function to display binding energy (Function 2) and the function to organize information in the form of log notes (Function 3). About Function 2, they expressed that the labor was reduced by annexing the binding energies of the constituent elements of the sample to the spectra. The student commented that the energy positions are quite easy to see because they appear together on the screen, and it means that users do not have to remember or check binding energies every time. Also, the participants gave the opinion that compiling information in the form of log notes saves time and reduces the risk of errors. the expert said that if the software shows us what information is actually required, it is easier for people who have no experience to understand what information is needed. The student also added that it would be great to be able to go with the flow and not make mistakes just by going with the flow.

# 6. Discussions

The previous chapter describes the experiments in which we evaluate *synapse* in terms of quantity and quality. In this chapter, I discuss the results of evaluation experiments and then explained limitations and future work.

## 6.1. Discussion of SNR determination

As shown in the interview survey (Section 3.1), it is not easy for novice users to determine the necessary and sufficient SNR. Therefore, we formulated SNR using Equation 1 and visualized it as a guideline for determining the necessary and sufficient SNR.

The results of experiments described in Section 5.1 suggest that the definition in Equation 1 is valid and satisfies all three requirements I explained in Section 4.2: dimensionless, able to be compared across spectra, and monotonically increasing according to the number of iterations. Given the Equation 1, it is clear that SNR is dimensionless. Also, as shown in Figure 2(a) and 2(c), the SNR of Sr 3d spectra is obviously larger than that of Sr 4d spectra, and this intuition is consistent with the calculation. Finally, given the spectra shown in Figs. 2(b) and 2(d), calculated SNRs are almost monotonically increasing. Thus, we could conclude that the formula of SNR we suggest can be a great indicator when users decide when to stop the iterations.

Furthermore, the results of the user study reveal that calculating and visualizing SNRs helps novice users of photoemission spectroscopy. This means that the two findings do not only prove how valid the definition of SNR is but also how effective defining SNR itself is. In the past, novice users had to ask users with experience for help when they determine the SNR, but from now on, veteran users can leave the measurement, for example, by saying, "Keep it running until it reaches 40 dB". Certainly, further discussion and consensus-building are necessary regarding the validity of Equation 1 and the type and parameters of the filter to be used, but SNR visualization is considered to be a sufficient solution to the issue of photoemission spectroscopy.

**6.2. Limitations & future work**

In this section, we provide five limitations in this study: filters, data sets, data loading, and application to other measurements.

The selection of filters can be a limitation. Savitzky-Golay filter was used in this study. However, there are other noise-reducing filters other than the Savitzky-Golay filter, such as moving average filters and simple low-pass filters. There seems to be room for discussion regarding the type of filter and parameters used to calculate SNR; if SNR is not standardized and there are multiple definitions, there is a risk of further confusion for non-skilled users.

Also, the data sets we used in this study are limited in number and do not cover all cases. In photoemission spectroscopy, there is no common database that stores experimental data, which limits into data our team has and prevents us from freely debugging and validating the algorithms. Therefore, for future algorithm development, it is necessary to construct a common data set for photoemission spectroscopy.

The third limitation is how to load the data. The current version of *synapse* requires data files to be read one by one. As it is, it is not easy to determine SNRs in real-time. Ideally, the file should be monitored for changes and automatically updated. The file monitoring function is relatively easy to implement and will be incorporated into the application in the future.

Lastly, in this study, we focus on ordinary photoemission spectroscopy. As an option, we would like to expand *synapse* to angle-resolved photoemission spectroscopy (ARPES), which is becoming more popular. Indeed, the participants of the user study recommend the expansion of the target.

**7. Conclusion**

In this study, we proposed a measurement and analysis support system for users who are not familiar with photoemission spectroscopy. First, we conducted interviews and found the issues of peak identification and SNR determination in the experimental process of photoemission spectroscopy. In order to solve these problems, we have developed *synapse* with the following three functions: function to quantitatively calculate and visualize SNRs, function to record the binding energies of the constituent elements of a sample, and function to manage information in the form of log notes. Finally, the application was evaluated quantitatively and qualitatively, and it was found that SNR could be determined easily enough, although there is room for improvement to assist in peak identification.

In the future, we aim to solve the issues described in section 6.2 and build an application that is easier to use for novice users of photoemission spectroscopy.

**Appendix A: Architecture**

*synapse* is built on Electron, a JavaScript package to build cross-platform desktop apps. Using Electron, we can easily build native desktop apps instead of web-based apps activating on browsers, which avoids users to start apps by the command-line interfaces. In this appendix section, I explained Electron and its multi-process model.

**A.1. An overview of Electron**

Electron[7] can be characterized as three points: compatibility, web-based stacks, and use cases. Electron enables developers to readily build native apps compatible with the various operating systems: Mac, Windows, and Linux. Also, Electron can be used with HTML, CSS, and JavaScript, which are the stacks used in the development of webbased apps (Surely, we can use TypeScript, React, and any other packages essential to modern frontend development). Thus, Electron is trusted by VSCode, WhatsApp, Slack, Figma, Twitch, and so on.

---

[7] https://www.electronjs.org/ja/docs/latest

**A.2. A Multi-process model of Electron**

Electron applications have the multi-process model, where developers can use the main process and the renderer processes. In a short, the main process orchestrates multiple renderer processes, which are responsible for the user interface. The main process is unique to the Electron app and plays a role in activating the Electron app, managing renderer processes (app windows), and corresponding with operating systems. On the other hand, the renderer processes play a role in rendering interface, behaving according to web standards. Thus, we can easily and quickly migrate from web-based apps to native apps. The main process communicates with the renderer processes via inter-process communication (IPC). Using inter-process communication, the renderer processes obtain access to operating systems such as file systems.

The renderer processes act like browser tabs, so developers can obtain benefits from numerous JavaScript packages including React, Next.js, TypeScript, and MUI. In that respect, web-based app development is not different from native app development. If we want to move from web-based apps to native apps, all we have to do is to transplant client parts to renderer processes and server parts to the main process.

**Acknowledgements**

We would like to thank Prof. A. Sekiyama, Prof. T. Yoshida, Prof. I. Matsuda, Prof. K. Shimada, and all the participants of the interview survey and user study for their cooperation. Through discussion, we can clarify our research's orientation. Also, we really appreciate our lab members. They gave us a lot of advice and feedback, which greatly helped us. This work was partially supported by the Spintronics Research Network of Japan (Spin-RNJ).